\documentstyle[12pt,twoside,epsfig,subfigure]{paper2}
  \newcommand{\ccaption}[2]{
    \begin{center}
    \parbox{0.85\textwidth}{
      \caption[#1]{\small{\it{#2}}}
      }
    \end{center}
    }
\begin{document}            
\newcommand\sss{\scriptscriptstyle}
\newcommand\mug{\mu_\gamma}
\newcommand\mue{\mu_e}
\newcommand\muf{\mu_{\sss F}}
\newcommand\mur{\mu_{\sss R}}
\newcommand\muo{\mu_0}
\newcommand\me{m_e}
\newcommand\as{\alpha_{\sss S}}         
\newcommand\ep{\epsilon}
\newcommand\aem{\alpha_{\rm em}}
\newcommand\refq[1]{$^{[#1]}$}
\newcommand\avr[1]{\left\langle #1 \right\rangle}
\newcommand\lambdamsb{\Lambda_5^{\rm \sss \overline{MS}}}
\newcommand\qqb{{q\overline{q}}}
\newcommand\qb{\overline{q}}
\newcommand\MSB{{\rm \overline{MS}}}
\newcommand\DIG{{\rm DIS}_\gamma}
\renewcommand\topfraction{1}       
\renewcommand\bottomfraction{1}    
\renewcommand\textfraction{0}      
\setcounter{topnumber}{5}          
\setcounter{bottomnumber}{5}       
\setcounter{totalnumber}{5}        
\setcounter{dbltopnumber}{2}       
\newsavebox\tmpfig
\newcommand\settmpfig[1]{\sbox{\tmpfig}{\mbox{\ref{#1}}}}
\begin{titlepage}
\nopagebreak
\vspace*{-1in}
{\leftskip 11cm
\normalsize
\noindent   
\newline
CERN-TH.7527/94 \newline
GEF-TH-9/1994 \newline
IFUM 487/FT

}               
\vfill
\begin{center}
{\Large \bf \sc
Total Cross Sections for}

{\Large \bf \sc Heavy Flavour Production at HERA}
\vfill
{\bf Stefano Frixione}
\vskip .3cm
{Dip. di Fisica, Universit\`a di Genova, and INFN, Sezione di Genova, 
Genoa, Italy}\\
\vskip .6cm                                               
{\bf Michelangelo L. Mangano\footnotemark}
\footnotetext{Address after Feb. 1$^{\rm st}$ 1995:
CERN TH-Division, CH-1211 Geneva 23, Switzerland}
\vskip .3cm
{INFN, Scuola Normale Superiore and Dipartimento di Fisica, Pisa, Italy}\\
\vskip .6cm                                               
{\bf Paolo Nason\footnotemark}
\footnotetext{On leave of absence from INFN, Sezione di Milano, Milan, Italy.}
and
{\bf Giovanni Ridolfi\footnotemark}
\footnotetext{On leave of absence from INFN, Sezione di Genova, Genoa, Italy.}
\vskip .3cm
{CERN TH-Division, CH-1211 Geneva 23, Switzerland}
\end{center}
\vfill
\nopagebreak
\begin{abstract}
{\small
We compute total cross sections for charm and bottom photoproduction at HERA
energies, and discuss the relevant theoretical uncertainties. In particular we
discuss the problems arising from the small-$x$ region, the uncertainties in
the gluon parton density, and the uncertainties in the hadronic component of
the cross section. Total electroproduction cross sections, calculated in the
Weizs\"acker-Williams approximation, are also given. 
}
\end{abstract}
\vfill
CERN-TH.7527/94 \newline
December 1994    \hfill
\end{titlepage}

\section{Introduction}
Photoproduction of heavy flavours has been studied extensively in fixed target
experiments (ref.~[\ref{FixedTargetPhotoproduction}]). The measured total cross
sections for charm photoproduction turn out to be in reasonable agreement with
theoretical expectations (see ref.~[\ref{FMNRft}]). At HERA new opportunities
are available. On one hand, one can extend in energy previous charm production
studies, from center-of-mass energies around 30 GeV to energies up to 300 GeV.
Some preliminar results have already been presented
in ref.~[\ref{H1Charm}].
Small-$x$ effects in these high energy regimes may not be negligible for charm
production,  and it will be interesting to see whether they are important at
HERA. Bottom production will also be observable at HERA. The larger value of
the bottom quark mass makes QCD predictions more reliable. Furthermore, bottom
hadroproduction has been extensively studied at hadron colliders, and the
comparison with QCD predictions, although qualitatively good, presents some
problems (see refs. [\ref{UA1},\ref{CDF},\ref{D0}]).
Theoretical arguments indicate that
perturbation theory should describe photoproduction of heavy flavours more
reliably than hadroproduction. This is also supported by experimental evidence
in fixed-target charm production. HERA has therefore the opportunity to clarify
some open questions in bottom production. 

In this paper, we discuss the total cross sections for charm and
bottom photoproduction, and for
electroproduction in the Weizs\"acker-Williams approximation,
in the HERA energy regime.
Our analysis is based on the next-to-leading order
calculation of heavy-quark photoproduction and hadroproduction cross sections
performed in refs.~[\ref{EllisNason},\ref{NDE}]. Some results in heavy-quark
electroproduction at HERA have already been presented in ref.~[\ref{FMNRhera}].
The computer program we used for this work was developed in
refs.~[\ref{MNR},\ref{FMNRphoto}], and is available upon request.
We will not consider electroproduction cross sections for large photon
virtuality, which has been recently discussed in ref.~[\ref{Neerven}].

Our work is organized as follows. In section~2 we discuss charm
photoproduction, focussing especially on the theoretical uncertainties which
are specific of HERA physics. In section~3 we give our results for bottom
photoproduction. In section~4 we present our results for charm and bottom
production in $ep$ collisions, and in section~5 we give our conclusions. 

\section{Charm production}
\begin{figure}[tbhp]
  \begin{center}
    \mbox{
      \epsfig{file=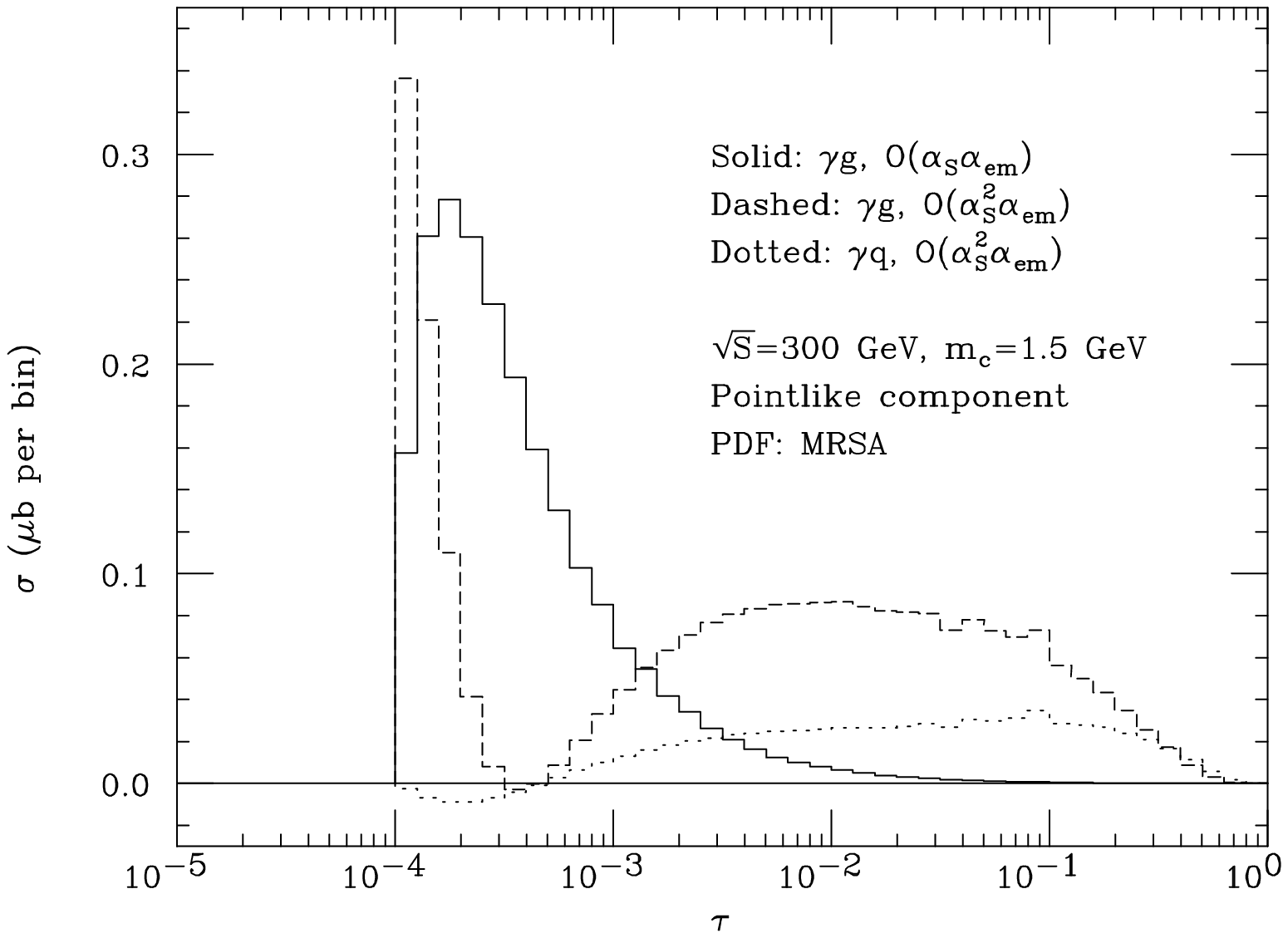,width=0.60\textwidth}
      }
  \ccaption{}{\label{tau_pnt_mrsa}
The ${\cal O}(\aem\as)$ $\gamma g$, ${\cal O}(\aem\as^2)$ $\gamma g$
and ${\cal O}(\aem\as^2)$ $\gamma q$ contributions to the total charm
photoproduction cross section, histogrammed as a function of 
$\tau=\hat{s}/S$. 
The proton factorization and renormalization scales have been taken
equal to $m_c$ in the short-distance cross sections.          }
  \end{center}
\end{figure}

In charm photoproduction at HERA,
because of the large center-of-mass energies available, the new problem
arises (which was not present in fixed target configurations) of the presence
of a large radiative effect, proportional to $\aem\as^2 \log(S/4
m_c^2)$, where $S$ is the squared center-of-mass energy of the
photon-hadron system, and $m_c$ is the charm quark mass. This
effect is the first manifestation of a whole tower of corrections behaving like
$\aem\as(\as \log(S/4 m_c^2))^n$, arising from
the small-$x$ region. In the high energy limit, $\as\log(S/4m_c^2)$ may become
of order 1, and one should then worry about resumming all these contributions.
A formalism for the resummation has been studied in refs.~[\ref{Smallx}]. In
the present work, we will try to assess the importance of these effects
at HERA. Following the work of ref.~[\ref{NDE}], in
fig.~\ref{tau_pnt_mrsa} we show various contributions to the total pointlike
photoproduction cross section, histogrammed as a function of the variable
$\tau$, defined as the ratio of the partonic versus the hadronic center-of-mass
squared energies $\hat{s}/S$. 

\begin{figure}[tbhp]
  \begin{center}
    \mbox{
      \epsfig{file=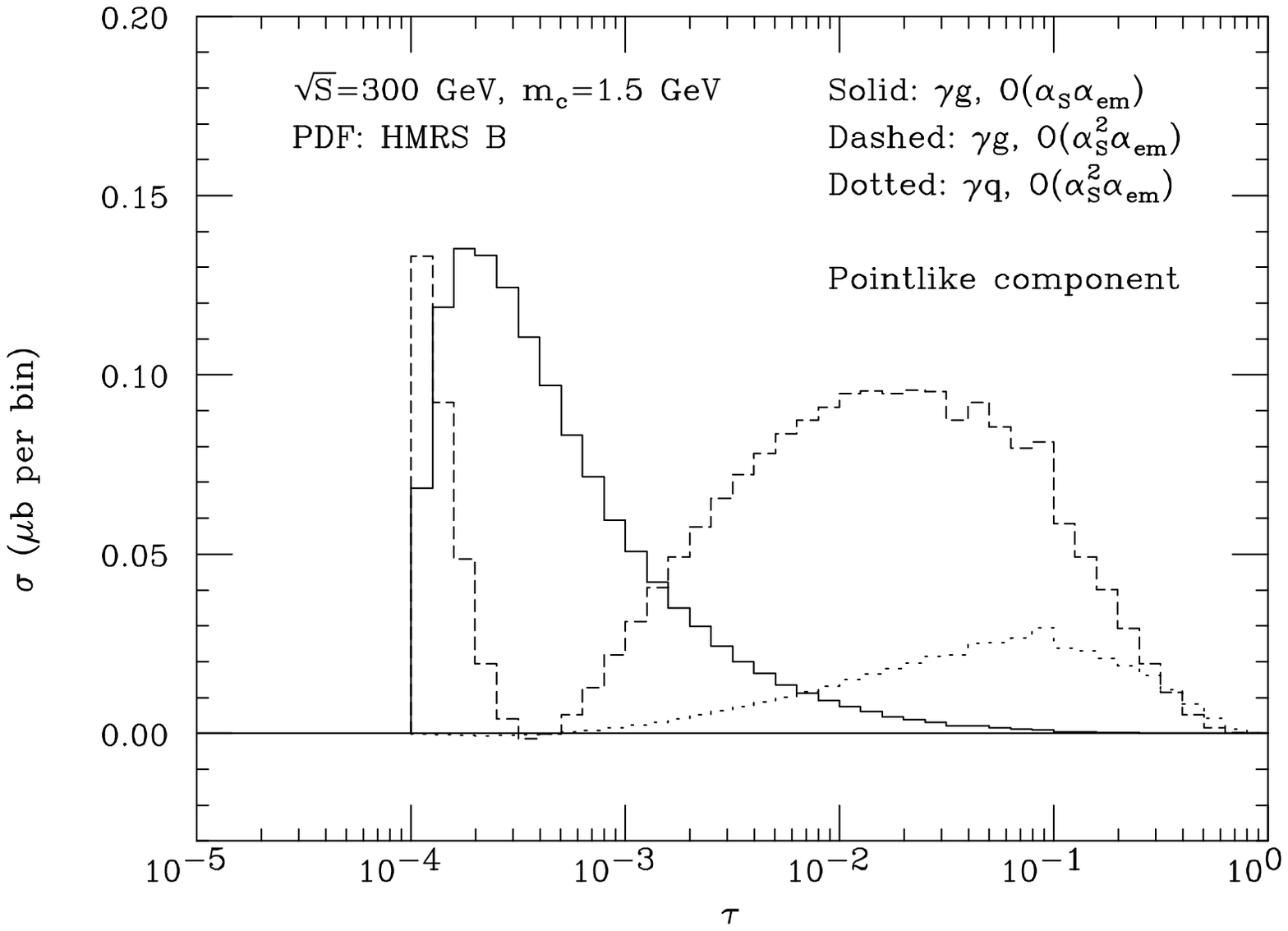,width=0.60\textwidth}
      }
\settmpfig{tau_pnt_mrsa}
  \ccaption{}{\label{tau_pnt_hmrs}
As in fig.~\box\tmpfig, using HMRS~B parton densities.
}
  \end{center}
\end{figure}

In the case of the pointlike component of photoproduction, this ratio
is simply equal to $x$, the fraction of the hadron
momentum carried by the parton. We see that in the
next-to-leading $\gamma g$ contribution there is a sizeable hump away from the
threshold region. The same behaviour is observed in the $\gamma q$
contribution. The hump arises from the fact that the next-to-leading parton
cross section goes to a constant for large $\hat{s}$ (see figs. 2 and 3 of
ref.~[\ref{EllisNason}]).
Therefore, for a hypothetical gluon density of the form $G(x)=1/x$,
the next-to-leading $\gamma g$ contribution to the total cross section would be
given by \beq \int_{4m_c^2/s}^1 \hat{\sigma}^{(1)}_{\gamma g}\; G(x)\,dx
=\int_{4m_c^2/s}^1 \hat{\sigma}^{(1)}_{\gamma g}\; d \log x. \eeq Observe that,
were it not for the fact that the gluon density has some extra damping factor
as $x\to 1$, the dashed histogram in fig.~\ref{tau_pnt_mrsa} would go to a
constant as $x\to 1$. In the very high energy limit, the hump will therefore
become a plateau, and its width would be roughly equal to $\log(S/4m_c^2)$. One
can show that this problem does not stop at this order: at order $\aem\as^n$
the partonic cross section grows like $(\log\hat{s}/4m_c^2)^{n-1}$, and this
would give a contribution to the total cross section raising like
$(\log(S/4m_c^2))^n$ with the energy. Notice that
fig.~\ref{tau_pnt_mrsa} was obtained using the MRSA parton densities
[\ref{MRSA}], which have a relatively singular gluon distribution at small $x$.
This type of parton density emphasizes the threshold region. We therefore
expect that when using parton densities with a more regular behaviour at small
$x$, the hump should be more pronounced. This is in fact the case, as shown in
fig.~\ref{tau_pnt_hmrs}, obtained using the parton densities HMRS~B\footnote{In
this case the parton density is frozen to its value at the minimum $Q^2$
allowed by the parametrization.} [\ref{HMRSB}] 

\begin{figure}[tbhp]
  \begin{center}
    \mbox{
      \epsfig{file=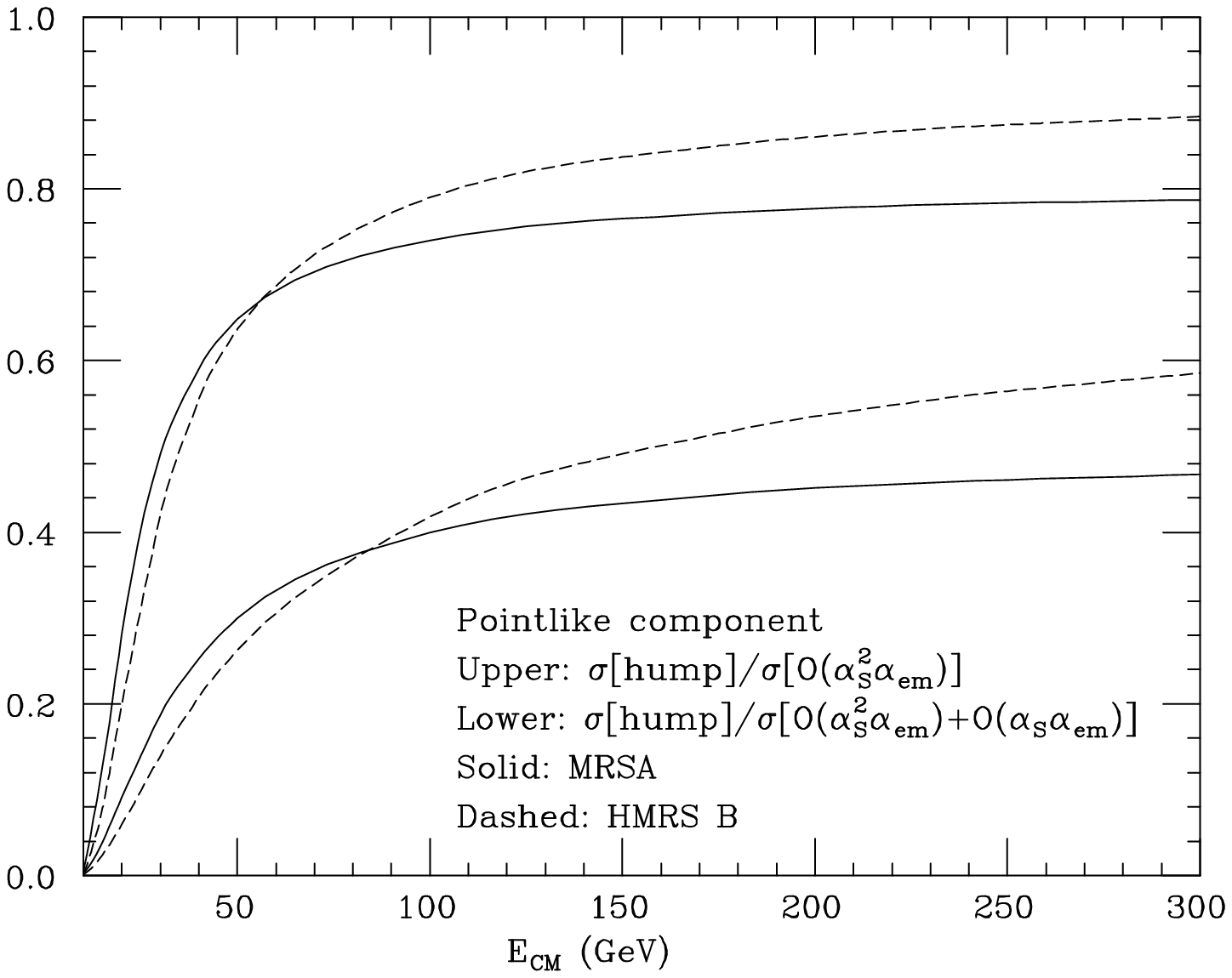,width=0.60\textwidth}
      }
  \ccaption{}{\label{hump_point}
The contribution of the hump to the total cross section relative
to the ${\cal O}(\as^2\aem)$ contribution (upper) and to the total
(lower) for the MRSA and HMRS~B proton parton densities.}
  \end{center}
\end{figure}

The size of the hump relative to the total cross section
can be taken as a measure of the importance of
the small-$x$ regime. In
fig.~\ref{hump_point} we plot the contribution to the total cross section
coming from the hump, relative to the total and to the ${\cal O}(\as^2\aem)$
contribution. The hump contribution is defined as the integral of the ${\cal
O}(\as^2\aem)$ term in $d\sigma/d\tau$, from its minimum point up to $\tau=1$. 
We are thus led to conclude that resummation effects could be important at the
typical center-of-mass energies explored by HERA. Furthermore, flatter parton
densities lead to larger small-$x$ effects. A similar pattern was found in the
context of hadroproduction of bottom quarks, in ref.~[\ref{CollinsEllis}],
where the all-order resummation of small-$x$ effects was actually performed for
$b$ production at the Tevatron. Lacking a calculation of the effects of
resummation in photoproduction, we will simply assume that it may give
an enhancement equal to the square of the size of the hump relative to the
total cross section, a pattern suggested by the results of
ref.~[\ref{CollinsEllis}]. From fig.~\ref{hump_point} we can then conclude that
our calculation may underestimate the total charm cross section by 20\% to
40\%, depending upon the parton density choice. 

Sizeable small-$x$ effects are also present in the hadronic component of the
charm cross section. In this case, we find that the relative importance of
the hump contribution varies between 20\% and 45\% when the center-of-mass
energy changes from 50 to 300 GeV (this result was obtained using the MRSA and
GRV--HO [\ref{GRV}] parton densities in the proton and the photon respectively).
The corresponding uncertainty on the final result is considerably smaller than
other uncertainties. 

Other important inputs in the computation of the total cross
section are the value of $\lambdamsb$ and the gluon density in the
proton. The world average for $\as(M_Z^2)$ taken from the Particle Data Book,
ref.~[\ref{PDB94}], is $\as(M_Z^2)=0.117\pm 0.005$, corresponding to
$\lambdamsb=215+65-55\,$MeV (the value of $\lambdamsb$ quoted in
ref.~[\ref{PDB94}] differs from ours because it actually refers to the
3-loop formula for $\as$, while we are quoting the 2-loop value).
On the other hand, the preferred value of
$\lambdamsb$ used in the latest parton density parametrizations of the CTEQ
group is 158~MeV (see ref.~[\ref{CTEQ3}]), while the MRSA parametrization uses
151~MeV (see ref.~[\ref{MRSA}]). These values are below the lowest
extreme of the
world average range. This is due to the well known fact that the determinations
of $\as$ from $e^+e^-$ data tend to be higher than those from deep-inelastic
scattering.
In order to assess the error in our predictions coming from the
uncertainties on $\lambdamsb$, we will therefore consider also
the parton density
set CTEQ2ML, which was fitted with $\lambdamsb=220\;$MeV, a value
close to the world average.
In summary, our range for $\lambdamsb$ is 151~MeV$ < \lambdamsb < 220\;$MeV,
which does not cover the full uncertainty on the present knowledge of $\as$.
We are forced to do so, since larger values of $\lambdamsb$ are too far
from deep inelastic scattering determinations, and it would be
difficult to assess their correlation with the gluon density in the proton.

\begin{figure}[tbhp]
  \begin{center}
    \mbox{
      \epsfig{file=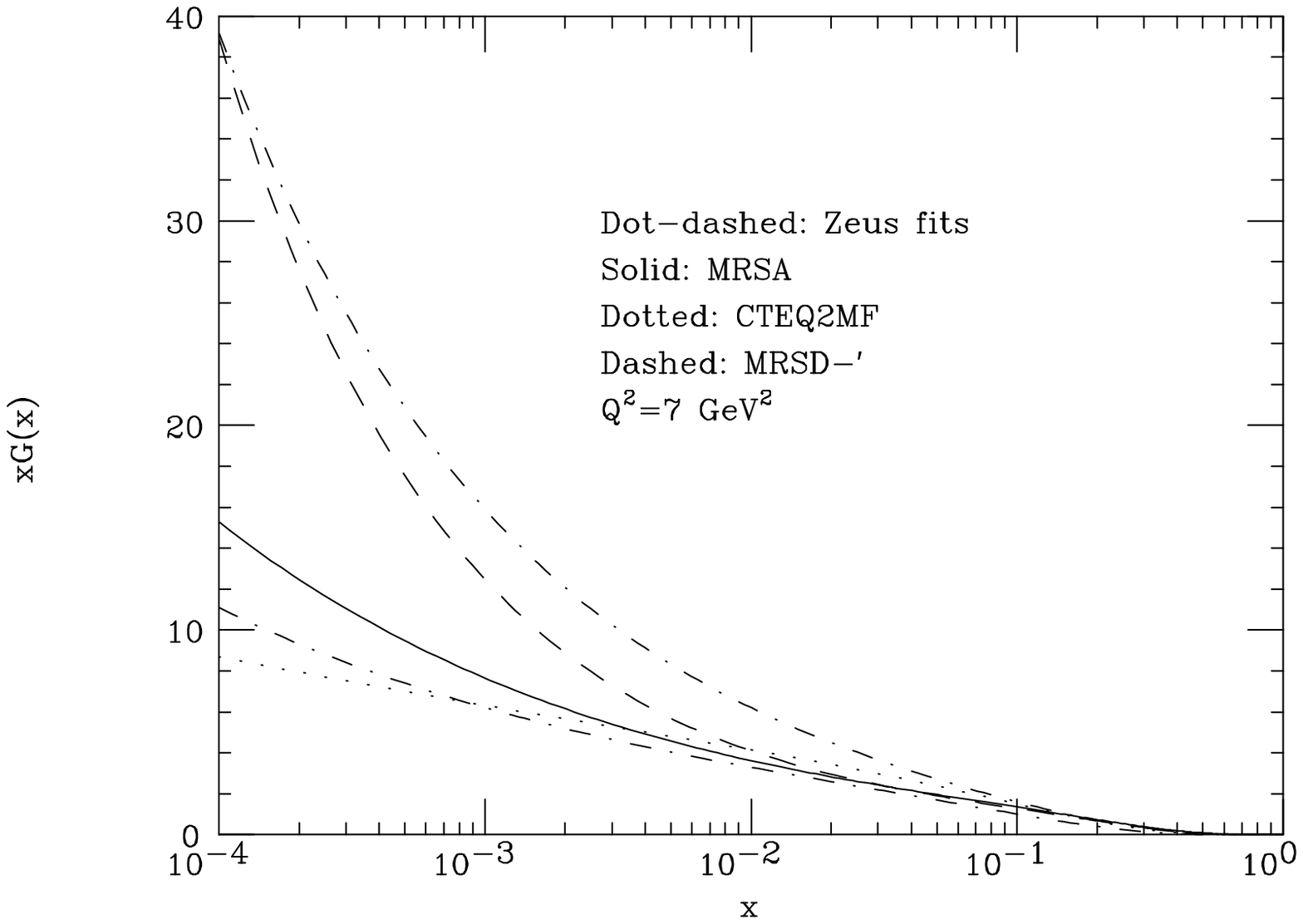,width=0.60\textwidth}
      }
 \ccaption{}{\label{gluons_at_small_x}
 The gluon densities in MRSA, MRSD--$^\prime$ and CTEQ2MF, together
 with the upper and lower parametrization of the ZEUS measurement.
}
  \end{center}
\end{figure}

Several uncertainties come into play in the extraction of the
gluon density. Its normalization is strongly constrained by the momentum sum
rule, while its shape, at relatively moderate value of $x$, is constrained by
direct photon production data. Deep inelastic scattering data constrain the
gluon density for relatively small values of $x$, where it affects the
evolution of the singlet parton density. At very small values of $x$, the gluon
density is in fact not yet known with precision. Both the CTEQ and MRS groups
assume that its small-$x$ behaviour is the same as that of $F_2$, which is
measured at HERA down to values of $x$ of the order of $10^{-4}$. On the other
hand, a measurement of the gluon density at $Q^2=7\;$GeV$^2$ has recently been
performed by the ZEUS collaboration\refq{\ref{Zeusgluon}}. We report in
fig.~\ref{gluons_at_small_x} the gluon densities given by the sets MRSA,
MRSD--$^\prime$ and CTEQ2MF, together with the two extreme small-$x$
parametrizations obtained by the ZEUS collaboration. 
The sets MRSD--$^\prime$ and CTEQ2MF differ
essentially in the assumed small-$x$ behaviour.  They are consistent with the
two extreme parametrizations of the ZEUS data, and we will therefore use them
as the two extreme possibilities.

\begin{figure}[tbhp]
\begin{center}
    \mbox{
      \epsfig{file=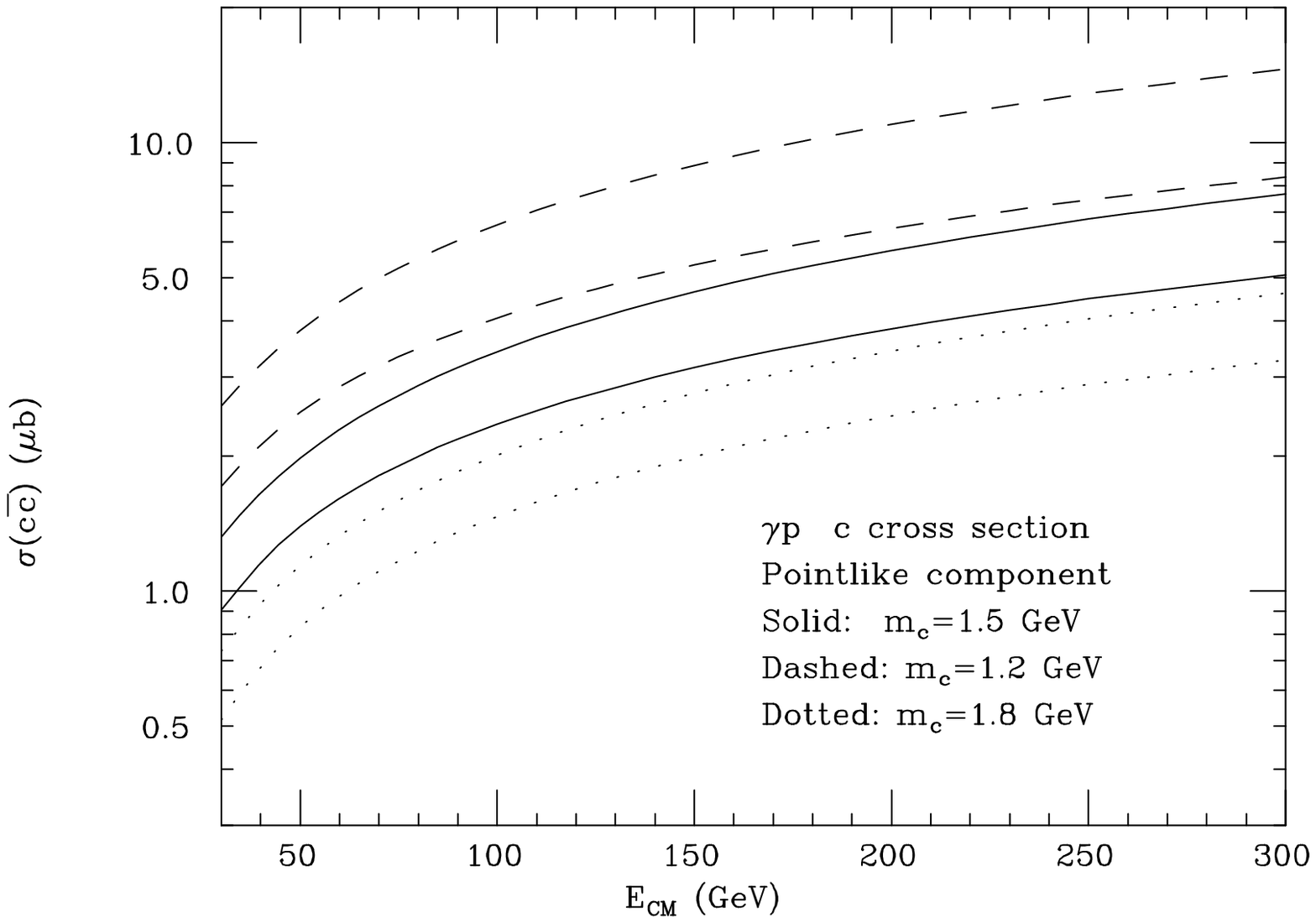,width=0.60\textwidth}
      }
 \ccaption{}{\label{band_c_mrsa}
 Pointlike component of the total charm cross section versus the
 center-of-mass energy in $\gamma p$ collisions. The bands
 are obtained by varying the renormalization
 scale between $m_c/2$ and $2 m_c$. The proton parton density set is MRSA.
}
  \end{center}
\end{figure}

\begin{figure}[tbhp]
  \begin{center}
    \mbox{
      \epsfig{file=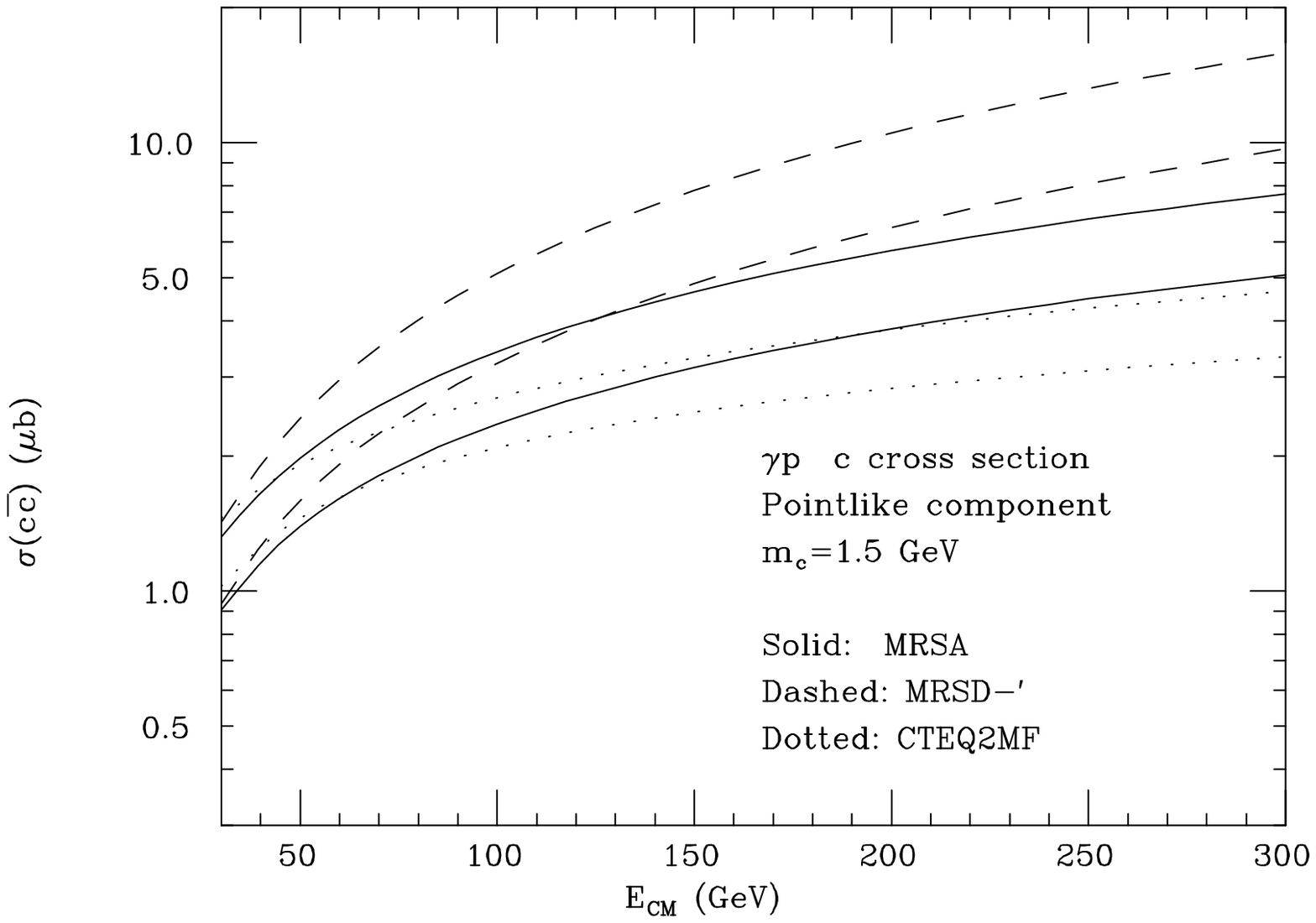,width=0.60\textwidth}
      }
 \ccaption{}{\label{band_c_lx}
 Pointlike component of the total charm cross section versus the
 center-of-mass energy in $\gamma p$ collisions for
 the three sets of parton densities MRSA, MRSD--$^\prime$ and CTEQ2MF. 
 The bands
 are obtained by varying the renormalization
 scale between $m_c/2$ and $2 m_c$.
}
  \end{center}
\end{figure}

Our results for the pointlike component of
the charm cross section are shown in table~\ref{c_point_tab} and in
figs.~\ref{band_c_mrsa} and \ref{band_c_lx}. The strong sensitivity of the
cross section to the charm quark mass is quite apparent. 
\begin{table}
\begin{center}
\begin{tabular}{|l||c|c|c|c|c|c|c|c|c|c|} \hline
& \multicolumn{3}{c|}{$m_c=1.2$ GeV} 
& \multicolumn{3}{c|}{$m_c=1.5$ GeV} 
& \multicolumn{3}{c|}{$m_c=1.8$ GeV} 
\\ \hline
$E_{\sss {\rm CM}}$ \ \ \ $\mur$ & $m_c/2$ &$m_c$ & 2$m_c$ &
  $m_c/2$ &$m_c$ & 2$m_c$ &
  $m_c/2$ &$m_c$ & 2$m_c$ \\ \hline \hline 
\multicolumn{10}{|c|}{Proton PDF set MRSA}
\\ \hline
30 GeV& 2.582 & 2.104 & 1.712 & 1.321 & 1.090 & 0.907 & 0.735 & 0.611 & 0.515
\\ \hline
100 GeV&6.558 & 5.056 & 4.057 & 3.415 & 2.826 & 2.356 & 2.009 & 1.721 & 1.467 
\\ \hline
280 GeV&13.89 & 10.13 & 8.008 & 7.314 & 5.858 & 4.831 & 4.389 & 3.685 & 3.120 
\\ \hline\hline
\multicolumn{10}{|c|}{Proton PDF set CTEQ2MF}
\\ \hline
30 GeV& 2.651 & 2.264 & 1.885 & 1.455 & 1.216 & 1.023 & 0.827 & 0.688 & 0.583
\\ \hline
100 GeV&4.371 & 3.856 & 3.236 & 2.699 & 2.425 & 2.088 & 1.761 & 1.593 & 1.393 
\\ \hline
280 GeV&7.292 & 5.681 & 4.607 & 4.504 & 3.827 & 3.237 & 3.035 & 2.676 & 2.317 
\\ \hline\hline
\multicolumn{10}{|c|}{Proton PDF set MRSD$-^\prime$}
\\ \hline
30 GeV& 3.136 & 2.363 & 1.885 & 1.422 & 1.136 & 0.936 & 0.748 & 0.612 & 0.513
\\ \hline
100 GeV&11.75 & 8.149 & 6.347 & 5.106 & 3.951 & 3.220 & 2.690 & 2.199 & 1.843 
\\ \hline
280 GeV&34.16 & 23.06 & 17.81 & 14.79 & 11.17 & 9.033 & 7.794 & 6.247 & 5.196 
\\ \hline\hline
\multicolumn{10}{|c|}{Proton PDF set CTEQ2M}
\\ \hline
30 GeV& 2.696 & 2.182 & 1.785 & 1.341 & 1.107 & 0.926 & 0.734 & 0.612 & 0.519
\\ \hline
100 GeV&6.618 & 5.240 & 4.262 & 3.473 & 2.917 & 2.454 & 2.051 & 1.773 & 1.522 
\\ \hline
280 GeV&13.46 & 10.17 & 8.157 & 7.207 & 5.898 & 4.919 & 4.374 & 3.728 & 3.184 
\\ \hline\hline
\multicolumn{10}{|c|}{Proton PDF set CTEQ2ML}
\\ \hline
30 GeV& 3.500 & 2.435 & 1.829 & 1.705 & 1.242 & 0.974 & 0.910 & 0.688 & 0.553
\\ \hline
100 GeV&8.818 & 5.625 & 4.171 & 4.270 & 3.169 & 2.497 & 2.451 & 1.942 & 1.583 
\\ \hline
280 GeV&18.90 & 10.52 & 7.618 & 8.844 & 6.205 & 4.823 & 5.145 & 3.972 & 3.213 
\\ \hline
\end{tabular}
\ccaption{}{\label{c_point_tab}
Pointlike component of the total charm cross sections ($\mu$b) in $\gamma p$
collisions, for various input parameters, as indicated.
}
\end{center}
\end{table}
Scale sensitivity is also important, giving an uncertainty of a factor of 2 for
the low mass value. From fig.~\ref{band_c_lx} we also see that by comparing the
cross section measured at HERA with lower energy measurements it should be
possible to distinguish between different small-$x$ behaviour of the structure
functions. 

\begin{figure}[tbhp]
  \begin{center}
    \mbox{
      \epsfig{file=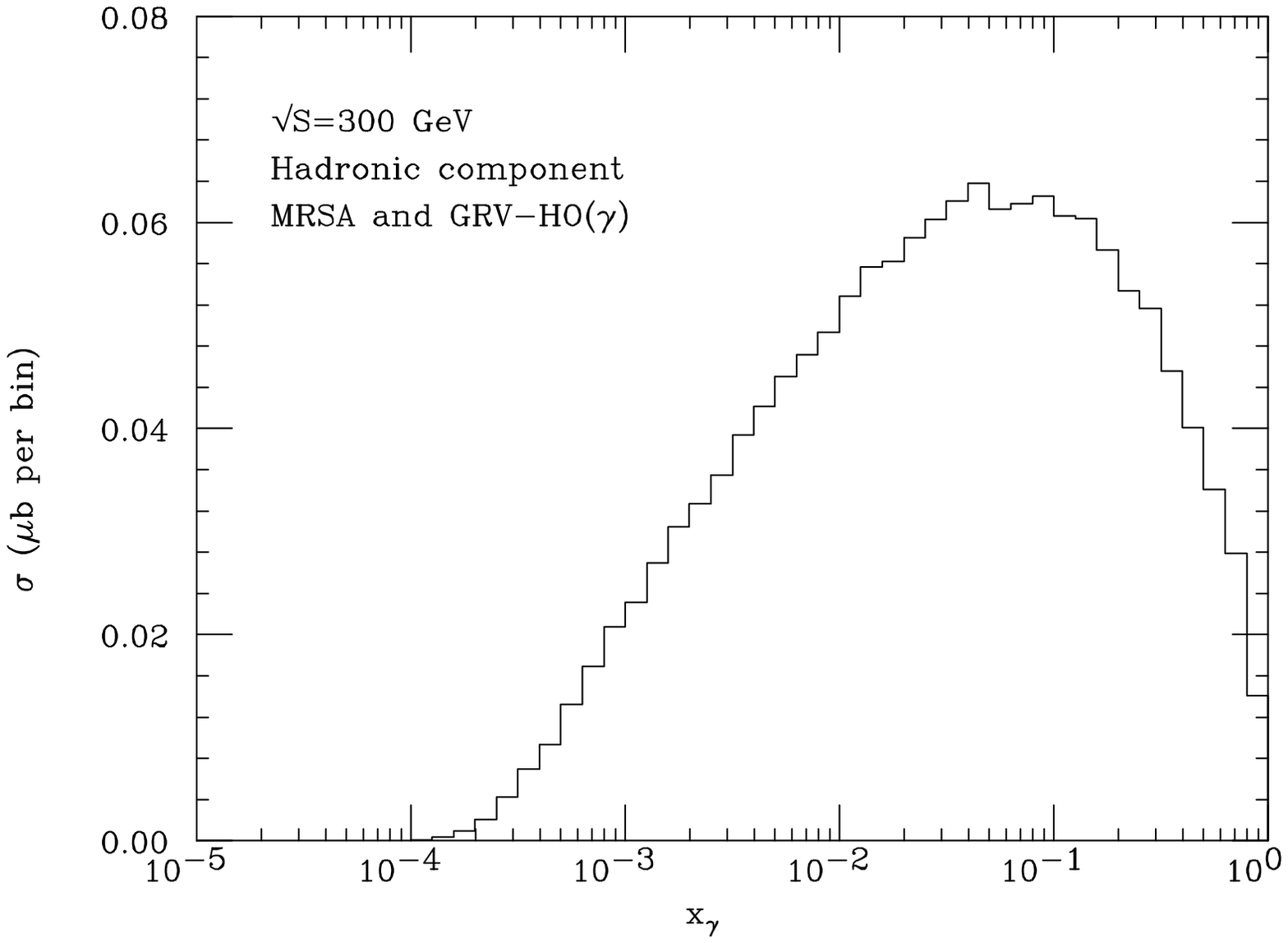,width=0.60\textwidth}
      }
  \ccaption{}{\label{SigmaVersusXgamma}
The $x_\gamma$ distribution of the hadronic component for charm
production in $\gamma p$ collisions.
}
  \end{center}
\end{figure}

\begin{figure}[tbhp]
  \begin{center}
    \mbox{
      \epsfig{file=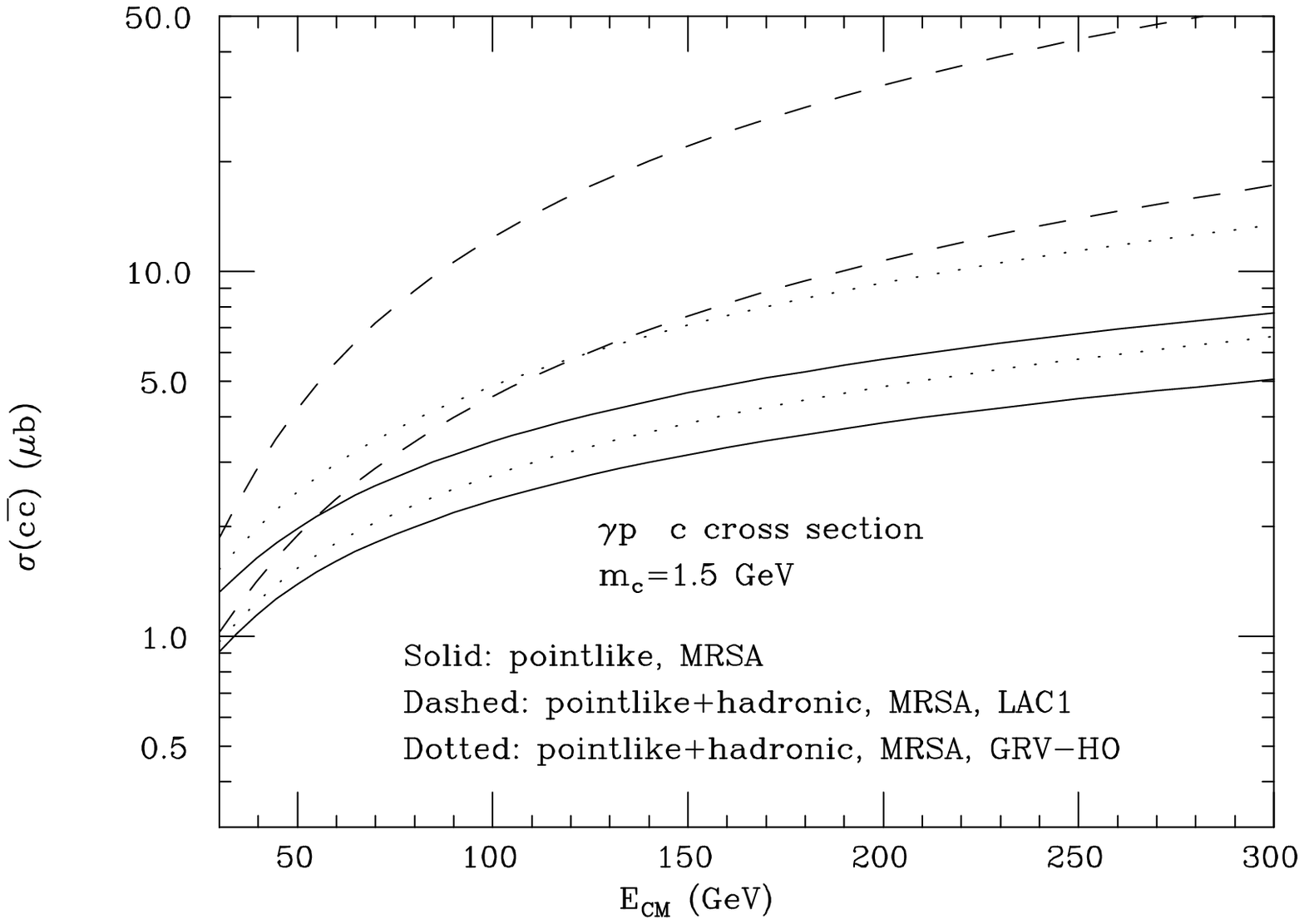,width=0.60\textwidth}
      }
 \ccaption{}{\label{band_c_had}
 Total charm cross section versus the
 center-of-mass energy in $\gamma p$  collisions, with and without 
 the inclusion of the hadronic component, computed with
 the two sets of photon parton densities LAC1 and GRV--HO.
 The bands are obtained by varying the renormalization
 scale between $m_c/2$ and $2 m_c$.
}
  \end{center}
\end{figure}

The hadronic component of the charm cross section is sensitive to the behaviour
of the gluon density in the photon at relatively small values of $x$, a region
not yet well explored experimentally. In fig.~\ref{SigmaVersusXgamma} we show
the hadronic component cross section histogrammed as a function of $x_\gamma$
(the partonic $x$ in the photon). 
We see that the dominant values of $x_\gamma$ are between $10^{-2}$ and
$10^{-1}$. Results from jet photoproduction at HERA\refq{\ref{Haas}} give some
indication on the gluon density in the photon, whose extraction depends upon
the value of the quark densities in the photon in a region where they are not
directly probed. These data, in spite of the uncertainties,
seem to indicate that the gluon density is
between the one in the GRV--HO set [\ref{GRV}] and the LAC1 set [\ref{LAC1}].
We note that the available experimental information on the gluon density in the
photon do not allow to distinguish among different next-to-leading order
parametrizations [\ref{NLOpdfph}] of the parton distributions.
Furthermore, the gluon density in the more recent leading order
parametrization of ref.~[\ref{Watanabe}] is within the range spanned
by the GRV--HO and the LAC1 sets. 
We will therefore take the sets GRV--HO and LAC1 as the two extremes.
For the proton
densities we will use the MRSA set. The relevant results are reported in
table~\ref{c_hadr_tab} and in fig.~\ref{band_c_had}. In
fig.~\ref{band_c_had} we plot the
cross section without hadronic component, and with
the hadronic component evaluated with the GRV--HO and LAC1 sets of photon parton
densities. We conclude that with the present uncertainty it is quite possible
that the hadronic component of the photon dominates the cross section at
typical HERA energies. The pointlike and hadronic components of the cross
section may be separated experimentally, at least in a first approximation,
since in the case of the hadronic component a sizeable fraction of the photon
momentum is lost into hadronic fragments. Therefore, at HERA it may be possible
to distinguish between the two sets of parton densities in the photon we have
considered.
\begin{table}
\begin{center}
\begin{tabular}{|l||c|c|c|c|c|c|c|c|c|c|} \hline
& \multicolumn{3}{c|}{$m_c=1.2$ GeV} 
& \multicolumn{3}{c|}{$m_c=1.5$ GeV} 
& \multicolumn{3}{c|}{$m_c=1.8$ GeV} 
\\ \hline
$E_{\sss {\rm CM}}$ \ \ \ $\mur$ & $m_c/2$ &$m_c$ & 2$m_c$ &
  $m_c/2$ &$m_c$ & 2$m_c$ &
  $m_c/2$ &$m_c$ & 2$m_c$ \\ \hline \hline 
\multicolumn{10}{|c|}{Photon PDF set GRV--HO}
\\ \hline
30 GeV& 0.793 & 0.297 & 0.165 & 0.206 & 0.099 & 0.060 & 0.072 & 0.040 & 0.026
\\ \hline
100 GeV&4.572 & 1.680 & 0.924 & 1.438 & 0.673 & 0.405 & 0.589 & 0.314 & 0.200
\\ \hline
280 GeV&15.42 & 5.492 & 2.989 & 5.296 & 2.435 & 1.454 & 2.348 & 1.239 & 0.784
\\ \hline\hline
\multicolumn{10}{|c|}{Photon PDF set LAC1}
\\ \hline
30 GeV& 3.350 & 0.995 & 0.503 & 0.540 & 0.213 & 0.119 & 0.125 & 0.059 & 0.036
\\ \hline
100 GeV&35.05 & 11.69 & 6.207 & 8.975 & 3.794 & 2.187 & 3.052 & 1.469 & 0.893
\\ \hline
280 GeV&140.3 & 48.77 & 26.30 & 42.45 & 18.84 & 11.09 & 16.78 & 8.494 & 5.283
\\ \hline
\end{tabular}
\ccaption{}{\label{c_hadr_tab}
Hadronic component of the total charm cross sections ($\mu$b) in $\gamma p$
collisions, for various input parameters, as indicated. The proton parton
density set is MRSA.
}
\end{center}
\end{table}                                

\section{Bottom production}
In the case of bottom production, all the uncertainties we have
discussed for charm are strongly reduced. The small-$x$ problem is much less
dramatic. We have performed a study of the bottom cross section versus $\tau$,
analogous to the one presented for charm in the previous section, and found
that the size of the hump contribution is of the order of 20\% ot the total
cross section. The sensitivity of the pointlike cross
section to the other parameters is given in
table~\ref{b_point_tab}. Our central values of the cross section, in
the column marked with DEF, are obtained with the renormalization and
factorization scales set equal to $m_b$, and $m_b=4.75$~GeV.
\begin{table}
\begin{center}
\begin{tabular}{|l||c||c|c||c|c||c|c|} \hline
& \multicolumn{1}{c||}{} 
& \multicolumn{2}{c||}{$\mur$} 
& \multicolumn{2}{c||}{$\muf$} 
& \multicolumn{2} {c|}{$m_b$ (GeV)} 
\\ \hline
$E_{\sss {\rm CM}}$ & DEF &$m_b/2$ & 2$m_b$ & $m_b/2$ & 2$m_b$ & 4.5 & 5
\\ \hline \hline 
\multicolumn{8}{|c|}{Proton PDF set MRSA}
\\ \hline
100 GeV& 22.90 & 26.00 & 20.36 & 21.87 & 23.63 & 27.41 & 19.28
\\ \hline
280 GeV& 64.30 & 75.44 & 56.09 & 57.81 & 66.95 & 75.31 & 55.32
\\ \hline\hline
\multicolumn{8}{|c|}{Proton PDF set CTEQ2MF}
\\ \hline
100 GeV& 24.91 & 27.93 & 22.33 & 23.80 & 25.73 & 29.70 & 21.02 
\\ \hline
280 GeV& 62.91 & 73.08 & 55.24 & 56.27 & 65.80 & 72.76 & 54.71
\\ \hline\hline
\multicolumn{8}{|c|}{Proton PDF set MRSD$-^\prime$}
\\ \hline
100 GeV& 23.03 & 26.19 & 20.45 & 22.18 & 23.63 & 27.75 & 19.26
\\ \hline
280 GeV& 74.76 & 87.24 & 65.43 & 69.22 & 76.93 & 88.76 & 63.52
\\ \hline\hline
\multicolumn{8}{|c|}{Proton PDF set CTEQ2M}
\\ \hline
100 GeV& 22.96 & 25.92 & 20.50 & 22.04 & 23.54 & 27.50 & 19.31
\\ \hline
280 GeV& 65.33 & 75.96 & 57.32 & 59.43 & 67.51 & 76.40 & 56.23
\\ \hline\hline
\multicolumn{8}{|c|}{Proton PDF set CTEQ2ML}
\\ \hline
100 GeV& 25.07 & 29.53 & 21.76 & 23.38 & 25.92 & 30.03 & 21.07
\\ \hline
280 GeV& 71.04 & 86.84 & 60.42 & 61.74 & 74.39 & 82.99 & 61.20
\\ \hline
\end{tabular}
\ccaption{}{\label{b_point_tab}
Pointlike component of the total bottom cross sections ($n$b) in $\gamma p$
collisions, for various input parameters, as indicated.
}
\end{center}
\end{table}                                
If we vary all the parameters together in the direction that makes the cross
section larger or smaller we get a spread for the bottom cross section which
goes from 16~$n$b to 35~$n$b at a center of mass energy of 100 GeV,
and from 41~$n$b to 101~$n$b at 280 GeV. The results for the hadronic
component are collected in table~\ref{b_hadr_tab}. The default values
correspond to $\mur=\muf=\mug=m_b=4.75$~GeV. 
\begin{table}
\begin{center}
\begin{tabular}{|l||c||c|c||c|c||c|c||c|c|} \hline
& \multicolumn{1}{c||}{} 
& \multicolumn{2}{c||}{$\mur$} 
& \multicolumn{2}{c||}{$\muf$} 
& \multicolumn{2}{c||}{$\mug$} 
& \multicolumn{2}{c|}{$m_b$ (GeV)} 
\\ \hline
$E_{\sss {\rm CM}}$ & DEF &$m_b/2$ & 2$m_b$ &$m_b/2$ & 2$m_b$ 
& $m_b/2$ & 2$m_b$ & 4.5 & 5
\\ \hline \hline 
\multicolumn{10}{|c|}{Photon PDF set GRV--HO}
\\ \hline
100 GeV& 4.244 & 5.589 & 3.290 & 4.246 & 4.258 & 3.883 & 4.525 & 5.444 & 3.347 
\\ \hline
280 GeV& 27.63 & 38.67 & 20.74 & 25.68 & 29.13 & 25.39 & 28.68 & 34.29 & 22.48
\\ \hline\hline
\multicolumn{10}{|c|}{Photon PDF set LAC1}
\\ \hline
100 GeV& 6.693 & 9.264 & 5.053 & 7.157 & 6.394 & 7.161 & 6.348 & 9.309 & 4.890
\\ \hline
280 GeV& 98.45 & 137.2 & 74.05 & 96.07 & 100.6 & 106.4 & 92.18 & 128.7 & 76.15
\\ \hline
\end{tabular}
\ccaption{}{\label{b_hadr_tab}
Hadronic component of the total bottom cross sections ($n$b) in $\gamma p$
collisions, for various input parameters, as indicated. The proton
parton density set is MRSA.
}
\end{center}
\end{table}                                
We see that in this case the effect of the hadronic component is less dramatic.
According to the LAC1 set, the hadronic contribution could be a substantial
fraction of the total, while the GRV--HO set gives a much smaller result. 

\section{Electroproduction of heavy quarks}
We present here our predictions for total charm and bottom  cross
sections in $ep$ collisions. The total electroproduction cross section in the
Weizs\"acker-Williams approximation is given by 
\beq
\sigma_{ep}(S)=\int dx \, f^{(e)}_\gamma(x,\mu_{\sss WW})\sigma_{\gamma p}(xS).
\eeq
In the following, unless otherwise specified, we will refer to the case when no
anti-tag condition is imposed on the outgoing electron. The function $
f^{(e)}_\gamma $ is given in this case by 
\beq
f^{(e)}_\gamma(x,\mu_{\sss WW})=\frac{\aem}{2\pi}\frac{1+(1-x)^2}{x}
\log\frac{\mu_{\sss WW}^2(1-x)}{m_e^2 x^2}.
\eeq
The scale $\mu_{\sss WW}$, according to ref. ~[\ref{FMNRww}], will be taken
equal to the quark mass for the pointlike component. 

The choice of $\mu_{\sss WW}$ in the case of the hadronic component is a
debatable issue. Intuitive reasoning leads to conclude that this scale should
not be larger than a typical hadronic scale, because a strongly off-shell
photon would have smaller probability of mixing with hadronic
resonances.
Some authors (see refs.~[\ref{Uematsu},\ref{BorSch},\ref{DreGod}])
suggest instead that this scale should be related to the
hardness of the process, and should therefore be larger for bottom production
than for charm production. In practice, in the charm case, changing $\mu_{\sss
WW}$ from $1$ GeV up to the quark mass has only a few percent effect on the
size of the hadronic component for charm production, while for bottom
production the effect could be as large as 20\%. In both cases, as we will see,
these effects are, at present, much smaller than other sources of uncertainty.
In the following we will choose $\mu_{\sss WW}=1\;$GeV for charm and $\mu_{\sss
WW}=3\;$GeV for bottom. We also notice that, in the case when an anti-tagging
condition is imposed, this problem is not relevant at all. 

\begin{table}
\begin{center}
\begin{tabular}{|l||c|c|c|c|c|c|c|c|c|c|} \hline
& \multicolumn{3}{c|}{$m_c=1.2$ GeV} 
& \multicolumn{3}{c|}{$m_c=1.5$ GeV} 
& \multicolumn{3}{c|}{$m_c=1.8$ GeV} 
\\ \hline
$\phantom{CTEQ2}\mur$ & $m_c/2$ &$m_c$ & 2$m_c$ &
  $m_c/2$ &$m_c$ & 2$m_c$ &
  $m_c/2$ &$m_c$ & 2$m_c$ \\ \hline \hline 
MRSA & 1.478 & 1.113 & 0.888 & 0.749 & 0.605 & 0.500 & 0.431 & 0.361 & 0.305 
\\ \hline
CTEQ2MF& 1.130 & 0.923 & 0.761 & 0.630 & 0.535 & 0.453 & 0.385 & 0.334 & 0.288
\\ \hline
MRSD$-^\prime$& 2.472 & 1.713 & 1.333 & 1.087 & 0.833 & 0.677 & 0.574 
              & 0.463 & 0.387
\\ \hline
CTEQ2ML& 2.078 & 1.250 & 0.918 & 0.956 & 0.678 & 0.529 & 0.530 & 0.406 & 0.327
\\ \hline
\end{tabular}
\ccaption{}{\label{c_el_point}
Pointlike component of the total charm cross sections ($\mu$b) in $ep$
collisions, at $E_{\sss CM}=314$ GeV, for various input parameters, 
as indicated. 
}
\end{center}
\end{table}                                
We present in tables~\ref{c_el_point} and \ref{c_el_hadr} our predictions for
the total charm cross sections, for the pointlike and the hadronic component
respectively, at $E_{\sss CM}=314$ GeV. 
The proton and photon parton densities
are chosen as in the case of photoproduction. We do not consider here the
CTEQ2M set, which gives essentially the same results as our default set MRSA.
We observe that the most important source of uncertainty is the charm mass. The
ratio bewteen the extreme choices $m_c=1.2$ and 1.8 GeV can be as large as 4.3
for the pointlike component and 10 for the hadronic component, and becomes
larger for smaller values of the renormalization scale. The renormalization
scale variation has a strong effect on the hadronic component (a factor up to
5.5 for the lowest value of the charm mass), and quite a sizeable effect on the
pointlike component. As in the photoproduction  case, the differences in the
hadronic component evaluated with GRV--HO and LAC1 densities are
striking. For our central values of
$m_c=1.5$ GeV and $\mur=m_c$, the hadronic contribution obtained
with the GRV--HO set is 20\% of the pointlike component,
while the hadronic and pointlike contributions are comparable
when using the LAC1 set. 
\begin{table}
\begin{center}
\begin{tabular}{|l||c|c|c|c|c|c|c|c|c|c|} \hline
& \multicolumn{3}{c|}{$m_c=1.2$ GeV} 
& \multicolumn{3}{c|}{$m_c=1.5$ GeV} 
& \multicolumn{3}{c|}{$m_c=1.8$ GeV} 
\\ \hline
$\ \ \ \ \mur$ & $m_c/2$ &$m_c$ & 2$m_c$ &
  $m_c/2$ &$m_c$ & 2$m_c$ &
  $m_c/2$ &$m_c$ & 2$m_c$ \\ \hline \hline 
GRV--HO& 0.878 & 0.320 & 0.176 & 0.277 & 0.129 & 0.078 & 0.115 & 0.061 & 0.039
\\ \hline
LAC1& 6.697 & 2.258 & 1.204 & 1.796 & 0.774 & 0.450 & 0.646 & 0.319 & 0.196
\\ \hline
\end{tabular}
\ccaption{}{\label{c_el_hadr}
Hadronic component of the total charm cross sections ($\mu$b) in $ep$
collisions, at $E_{\sss CM}=314$ GeV, for various input parameters, 
as indicated. The MRSA set for the densities in the proton was used.
}
\end{center}
\end{table}                                

We also calculated charm electroproduction cross sections with the anti-tagging
condition imposed, as in ref.~[\ref{FMNRhera}]; the results can be
obtained from those presented in
tables~\ref{c_el_point} and \ref{c_el_hadr} by multiplying by a factor of
0.7, which is approximately independent of the parameters of the calculation. 

We have checked that varying the factorization scale of the photon in the
pointlike component has a very small effect on the results. 

In our analyses, we have kept the factorization scales $\muf$ and $\mug$ fixed,
since parton densities are usually available only for momentum scales above
$\sim 5$ GeV$^2$. There are actually two independent next-to-leading order
parametrizations of densities in the proton valid for $Q^2<m_c^2$: an extension
of the MRSA set~[\ref{MRSAmod}], which we will call MRSA modified, and the
GRV--HO set~[\ref{GRVprot}].
We have used these two parametrizations to perform a study of the
factorization scale dependence of the charm cross section.

The MRSA modified set is obtained by ordinary
Altarelli-Parisi evolution of the MRSA set, supplemented with some specific
assumptions on non-perturbative effects. On the other hand, the GRV--HO
set has no non-perturbative assumptions.
It is obtained by perturbative evolution, starting
from ``valence-like" inputs at $Q_0^2=0.3$ GeV$^2$. No attempt is made to mimic
the low-$Q^2$ data.
The parametrization of non-perturbative effects inserted in the MRSA set is,
strictly speaking, in contrast with the prescriptions of the factorization
theorem. Nevertheless, we verified that these phenomenologically motivated
inputs are numerically not important for factorization scales around the charm
mass and larger. As far as the GRV set is concerned, the validity of the
Altarelli-Parisi evolution equations and the perturbative expansion at the very
low momentum scale at which the evolution is started is questionable.
However, we believe that both sets are well suited for the purpose of
estimating the uncertainties arising from the factorization scale variation.

\begin{figure}[htbp]
  \begin{center}
    \mbox{
      \begin{tabular}[t]{cc}
        \subfigure[]{
          \epsfig{file=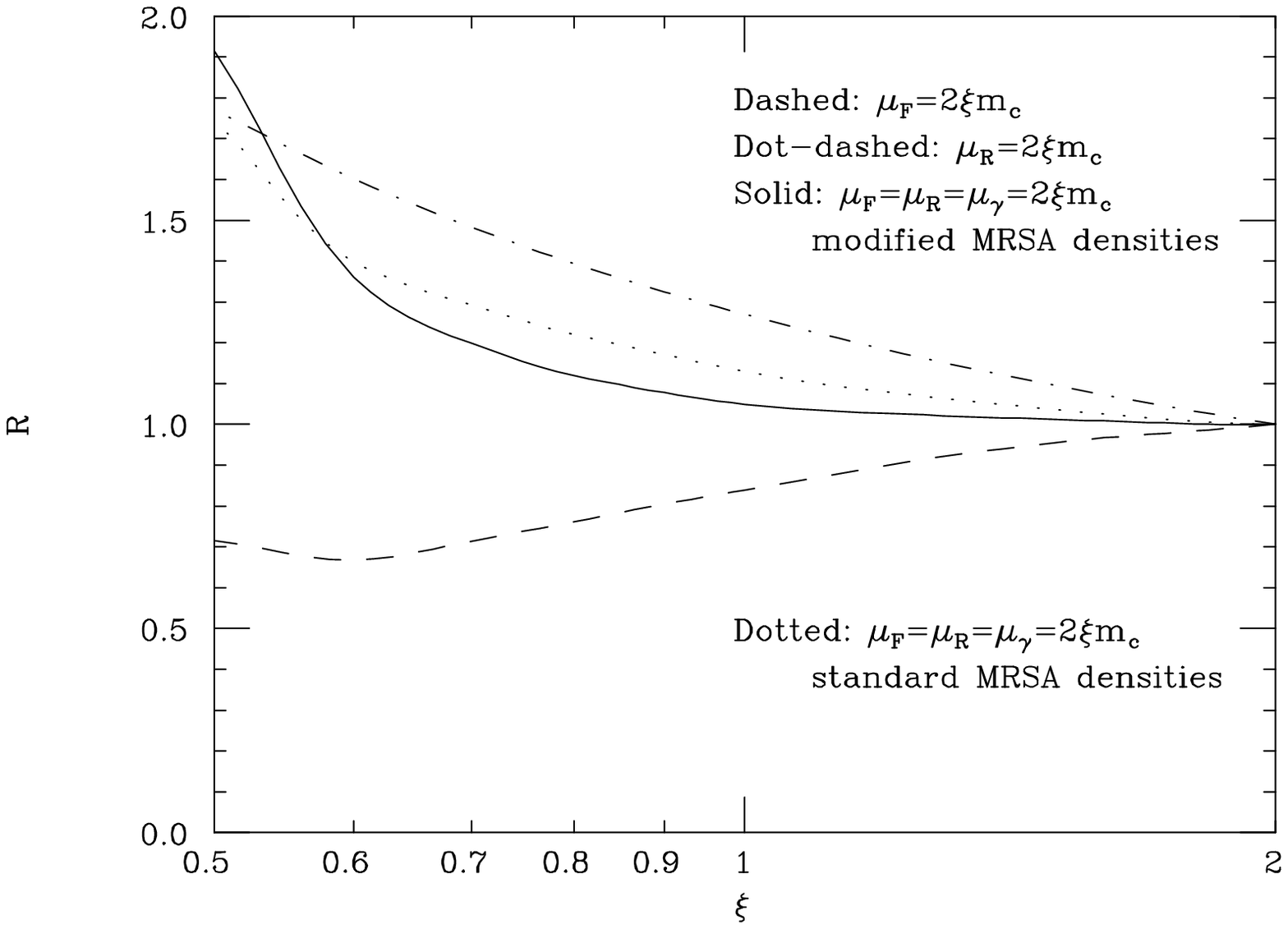,width=0.45\textwidth}
          } &
        \subfigure[]{
          \epsfig{file=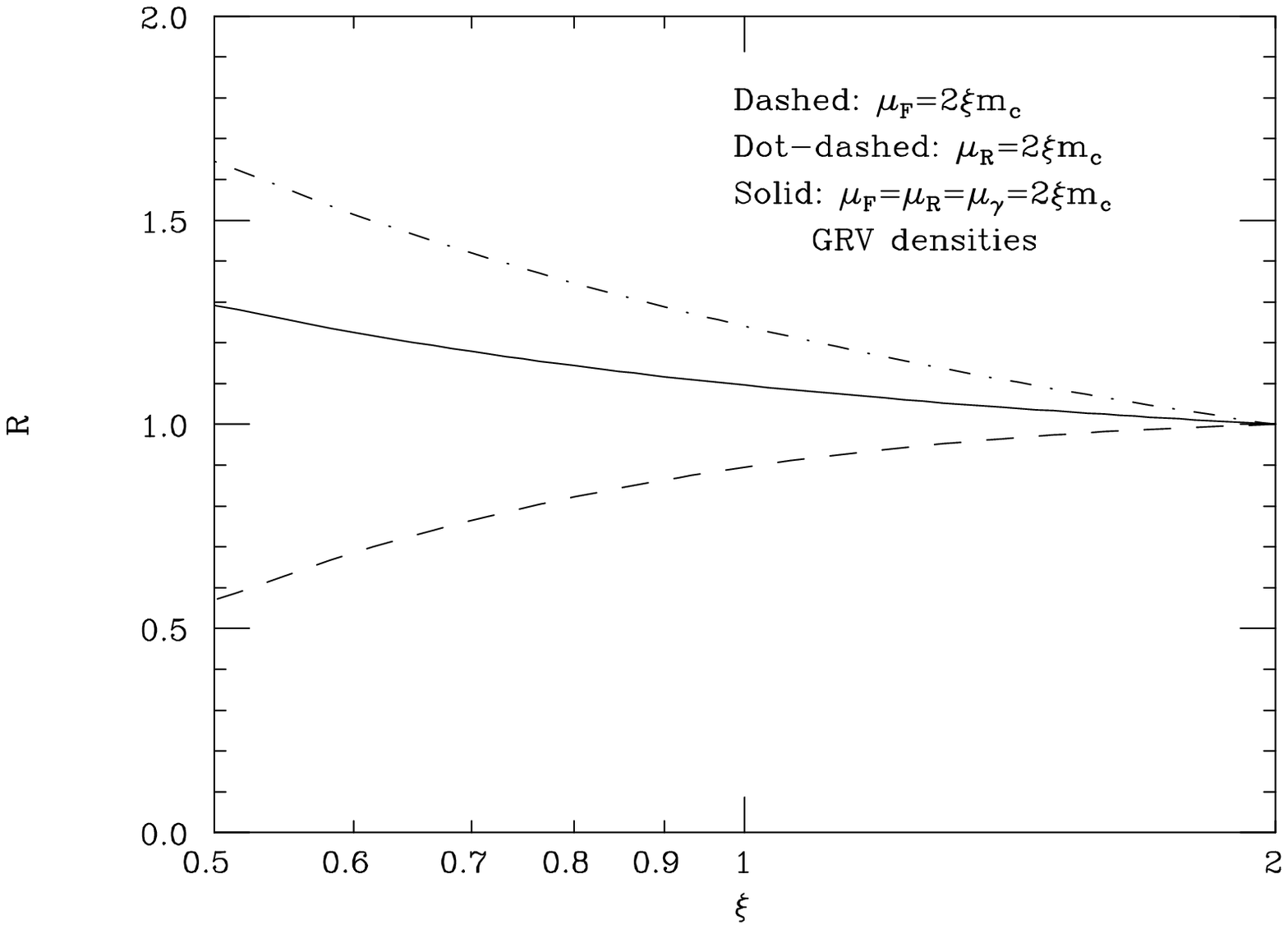,width=0.45\textwidth}
          }
      \end{tabular}
      }
  \end{center}
  \ccaption{}{
The ratio $R$, defined in the text, for different choices of factorization and
renormalization scales. MRSA (a) and GRV (b) densities in the proton are used.
  \label{lowq2muf}
    }
\end{figure}
In fig.~\ref{lowq2muf} we plot the ratio
\beq
R\,=\,\frac{\sigma(\mu=\xi m_c)}{\sigma(\mu=2m_c)}
\label{Rplot}
\eeq
as a function of $\xi$. The dot-dashed curve is obtained with
$\mur=\mu$, and keeping $\muf=\mug=2m_c$, the dashed curve has
$\muf=\mu$, $\mur=m_c$ and $\mug=2m_c$, and the solid curve has
$\muf=\mur=\mug=\mu$.
In the case of the MRSA parton densities, we also show (dotted
curve) the same quantity, for $\muf=\mur=\mug=\mu$, with the parametrization of
non-perturbative effects removed. In eq.~(\ref{Rplot}) $\sigma$ is the
full (poinlike plus hadronic)
cross section for an untagged electron and $m_c=1.5$ GeV. We use the
GRV--HO set
for the densities in the photon. When varying the hadronic factorization scale,
the cross section, in the case of the MRSA modified set, decreases approaching
the very low momentum region, where it increases again. This effect is
essentially due to target-mass corrections in the parton densities, which
sizeably harden the densities at low scales, and which are not present in the
GRV set. The solid line is the result for the simultaneous variation of all the
scales involved, including the photon factorization scale. We stress therefore
that in this case we are probing also photon densities at low momenta.
From this study we conclude that variation of the factorization scale
brings about an additional uncertainty of 40\% in the direction of
lower cross sections.

In tables~\ref{b_el_point} and \ref{b_el_hadr} we present the pointlike and
hadronic components of the total cross sections for bottom production at
$E_{\sss CM}=314$ GeV.
The defalut values correspond to all the scales equal to $m_b$,
and $m_b=4.75$~GeV. As expected, in this case the sensitivity to scale
variations is much smaller than in the charm case. The sensitivity to the
choice of parton densities is even less important than that arising from scale
variation. This is because the small-$x$ region, where the various sets mostly
differ, is not deeply probed in the case of bottom production. The differences
between results from various parton density sets come mainly from the different
values of $\Lambda$. 

\begin{table}
\begin{center}
\begin{tabular}{|l||c||c|c||c|c||c|c|} \hline
& \multicolumn{1}{c||}{} 
& \multicolumn{2}{c||}{$\mur$} 
& \multicolumn{2}{c||}{$\muf$} 
& \multicolumn{2} {c|}{$m_b$ (GeV)} 
\\ \hline
    & DEF &$m_b/2$ & 2$m_b$ & $m_b/2$ & 2$m_b$ & 4.5 & 5
\\ \hline \hline 
MRSA           & 4.640 & 5.332 & 4.098 & 4.391 & 4.773 & 5.541 & 3.917 
\\ \hline
CTEQ2MF        & 4.814 & 5.469 & 4.283 & 4.554 & 4.958 & 5.720 & 4.079 
\\ \hline
MRSD$-^\prime$ & 4.856 & 5.580 & 4.288 & 4.649 & 4.963 & 5.839 & 4.073
\\ \hline
CTEQ2ML        & 5.123 & 6.110 & 4.418 & 4.728 & 5.290 & 6.113 & 4.325     
\\ \hline
\end{tabular}
\ccaption{}{\label{b_el_point}
Pointlike component of the total bottom cross sections ($n$b) in $ep$
collisions, for various input parameters, as indicated, and
$E_{\rm CM}=314\;$GeV.
}
\end{center}
\end{table}                                

The hadronic component, displayed in table~\ref{b_el_hadr},
is at most 75\% of the pointlike component, even when the LAC1 set
is used. Again, this is related to the fact that the small-$x$ region, where
LAC1 gluon density is much larger than the GRV--HO one, is not so important in
this case.
\begin{table}
\begin{center}
\begin{tabular}{|l||c||c|c||c|c||c|c||c|c|} \hline
& \multicolumn{1}{c||}{} 
& \multicolumn{2}{c||}{$\mur$} 
& \multicolumn{2}{c||}{$\muf$} 
& \multicolumn{2}{c||}{$\mug$} 
& \multicolumn{2} {c|}{$m_b$ (GeV)} 
\\ \hline
    & DEF &$m_b/2$ & 2$m_b$ & $m_b/2$ & 2$m_b$ &$m_b/2$ & 2$m_b$ & 4.5 & 5
\\ \hline \hline 
GRV--HO & 1.115 & 1.510 & 0.851 & 1.077 & 1.146 & 1.018 & 1.177 & 1.408 & 0.892
\\ \hline
LAC1    & 2.842 & 3.947 & 2.142 & 2.861 & 2.840 & 3.067 & 2.675 & 3.792 & 2.157 
\\ \hline
\end{tabular}
\ccaption{}{\label{b_el_hadr}
Hadronic component of the total bottom cross sections ($n$b) in $ep$
collisions, for various input parameters, as indicated, and
$E_{\rm CM}=314\;$GeV. The proton parton density set is MRSA.
}
\end{center}
\end{table}                                
In the case of bottom, electroproduction cross sections with the anti-tagging
condition imposed on the scattered electron are approximately a factor 0.6
smaller than those presented in tables~\ref{b_el_point} and \ref{b_el_hadr},
independently of all other parameters. 

\section{Conclusions}
Charm and bottom photoproduction cross sections are theoretically in a better
shape than corresponding hadroproduction cross sections. We have presented
results which indicate that the small-$x$ effects should be under control at
HERA energies, even for charm production, and that a measurement of the
pointlike component of the charm photoproduction cross section at HERA could
help to discriminate among various possible small-$x$ behaviours of the gluon
densities in the proton and in the photon. 

We found that the most important source of uncertainty is coming from the
hadronic component of the cross section. If the gluon density in the photon is
as large as the LAC1 parametrization suggests, then the hadronic component
will be comparable or larger than the pointlike component for energies above
100 GeV for charm, and for energies of the order of 280 GeV for
bottom. Hopefully,
by looking at the structure of the events, experiments may be able
to separate the pointlike and hadronic components, and therefore
assess their relative magnitude. 

We also performed a study on total electroproduction cross sections in the
Weiz\-s\"acker-Williams approximation, including
recent parton density parametrizations, the effect of an anti-tagging condition
on the electron, and a more thorough illustration of the sensitivity to the
various physical parameters. We find results that are in substantial agreement
with earlier studies (ref.~[\ref{FMNRhera}]).

\section*{Acknowledgements} We would like to thank
M.~Fontannaz, K.~Hagiwara, G.~Iacobucci, R.~Prosi, J.~Roldan, A.~Staiano,
W.~J.~Stirling and M.~Tanaka for providing us with useful information. 

\clearpage
\begin{reflist}
\item \label{FixedTargetPhotoproduction}
    G. Bellini, proceedings of ``Les Rencontres de Physique de la Vallee
    d'Aoste'', La Thuile, Aosta Valley, March 6-12, 1994;\newline
    J.C.~Anjos et al., E691 Coll., \prl{62}{89}{513};\newline
    M.~P.~Alvarez et al., NA14/2 Coll., \zp{C60}{93}{53}.
\item \label{FMNRpheno}\label{FMNRft}
    S.~Frixione, M.~L.~Mangano, P.~Nason and G.~Ridolfi, preprint
    CERN-TH.7292/94, GEF-TH-4/1994, to appear in Nuclear Physics B.
\item \label{H1Charm}
    H1 Coll., preprint DESY-94-187, 1994.
\item\label{UA1}
    C. Albajar et al., UA1 Coll., \pl{B256}{91}{121}.
\item\label{CDF}
    K.~Byrum, CDF Coll., presented at the XXIV International Conference
    on High Energy Physics, Glasgow, Scotland, 20-27 July 1994,
    preprint FERMILAB-Conf-94/325-E.
\item\label{D0}
    D.~Hedin, D0 Coll., presented at the XXIV International Conference on
    High Energy Physics, Glasgow, Scotland, 20-27 July 1994
    preprint FERMILAB-Conf-94/415.
\item\label{Neerven}
    S.~Riemersma, J.~Smith and W.~L.~van Neerven, preprint 
    SMU-HEP 94-25, ITP-SB-94-59, INLO-PUB-16/94, hep-ph/9411431
\item\label{EllisNason}
    R.K.~Ellis and P.~Nason, \np{B312}{89}{551}.
\item \label{NDE}
   P.~Nason, S.~Dawson and R.~K.~Ellis, \np{B303}{88}{607}; {\bf B327}(1988)49.
\item \label{FMNRhera}
    S.~Frixione, M.~L.~Mangano, P.~Nason and G.~Ridolfi, \pl{B308}{93}{137}.
\item \label{MNR}
   M.L. Mangano, P. Nason and G. Ridolfi, \np{B373}{92}{295}.
\item \label{FMNRphoto}
   S.~Frixione, M.~L.~Mangano, P.~Nason and G.~Ridolfi, \np{B412}{94}{225}.
\item \label{Smallx}
   R.K. Ellis and D.A. Ross, \np{B345}{90}{79};\newline
   S. Catani, M. Ciafaloni and F. Hautmann, \pl{B242}{90}{97},
   \np{B366}{91}{135}, {\it Nucl. Phys. B (Proc. Suppl.)}
   {\bf 23B}(1991)328.
\item\label{MRSA}
   A.~D.~Martin, R.~G.~Roberts and W.~J.~Stirling, preprint RAL-94-055,
   DTP/94/34.
\item\label{HMRSB}
   P.~Harriman, A.~Martin, R.~Roberts and W.J.~Stirling,
   \pr{D37}{90}{798}.
\item\label{CollinsEllis}
   J.C. Collins and R.K. Ellis, \np{B360}{91}{3}.
\item \label{GRV}
    M.~Gl\"uck, E.~Reya and A.~Vogt, \pr{D46}{92}{1973}.
\item\label{PDB94}
    Review of Particle Properties, Phys. Rev. {\bf D50} Part I, p. 1297.
\item\label{CTEQ3}
   H.L. Lai et al., preprint MSU-HEP-41024, CTEQ-404, hep-ph-9410404.
\item\label{Zeusgluon}
    ZEUS Coll., preprint DESY-94-192, 1994.
\item \label{Haas}
    T. Haas, preprint DESY 94-160, Sept. 1994.
\item\label{LAC1}
    H.~Abramowicz, K.~Charchula and A.~Levy, \pl{269B}{91}{458}.
\item\label{NLOpdfph}
    L.~E.~Gordon and J.~K.~Storrow, \zp{C56}{92}{307};\\
    P.~Aurenche, M.~Fontannaz and J.-Ph.~Guillet, preprint ENSLAPP-A-435-93,
    LPTHE Orsay 93-37.
\item\label{Watanabe}
    K. Hagiwara, M. Tanaka and I. Watanabe, preprint KEK-TH-376,
    hep-ph/9406252.    
\item \label{FMNRww}
   S.~Frixione, M.~L.~Mangano, P.~Nason and G.~Ridolfi, 
   \pl{B319}{93}{339}.
\item \label{Uematsu}
   T.~Uematsu and I.~F.~Walsh, \np{B199}{82}{93}.
\item \label{BorSch}
   F.~Borzumati and G.~A.~Schuler, \zp{C58}{92}{139}.
\item \label{DreGod}
   M.~Drees and R.~M.~Godbole, \pr{D50}{94}{3124}.
\item \label{MRSAmod}
   A.~D.~Martin, R.~G.~Roberts and W.~J.~Stirling, preprint RAL-94-104,
   DTP/94/78.
\item \label{GRVprot}
   M.~Gl\"uck, E.~Reya and A.~Vogt, \zp{C53}{92}{127}.
\end{reflist}
\end{document}